\documentclass{aa}
\usepackage{graphicx}
\def\ssim{\setbox0=\hbox{$\sim$}%
\setbox1=\hbox{$<$}\dimen0=\ht1%
\advance\dimen0by-1.2pt\,\lower.6\dimen0%
\copy0\kern-\wd0\raise.4\dimen0\copy1 \,}

\def\gsim{\setbox0=\hbox{$\sim$}%
\setbox1=\hbox{$>$}\dimen0=\ht1%
\advance\dimen0by-1.2pt\,\lower.6\dimen0%
\copy0\kern-\wd0\raise.4\dimen0\copy1\,}

\def\lambdab{\lambda\mkern-9mu\lower1.2pt\hbox{$\mathchar'26$}}%

\begin{document}
   \title{The shape of  $\eta$ Carinae and LBV Nebulae}

 \author{A. Maeder
          \inst{1}
          \and
           V. Desjacques
           \inst{2}  
          }

   \offprints{A. Maeder}

   \institute{Geneva Observatory CH-1290 Sauverny, Switzerland
              email: Andre.Maeder@obs.unige.ch
         \and
             Geneva Observatory CH-1290 Sauverny, Switzerland
              email: Vincent.Desjacques@obs.unige.ch
             }

   \date{Received ...; accepted ...}

\abstract{Stellar winds emitted by
rotating  massive stars may show two main components:
firstly bipolar lobes  with  low density
and fast wind, produced by the higher  T$_{\mathrm{eff}}$ and 
gravity at the poles (``g$_{\mathrm{eff}}$--effect''); secondly,
an  equatorial disc  with a  slow  dense wind,
produced by the stronger opacities at the equator (``$\kappa$--effect'').
To see the possible role of this anisotropic wind
on the shape of LBV nebulae,
we calculate the distribution of the ejected matter in 2 simplified cases:
1) A  brief shell ejection. We find that prolate and 
peanut--shaped hollow nebulae  naturally form  due to the
g$_{\mathrm{eff}}$--effect in rotating stars.
2) A constant  wind for a long time. 
This produces prolate filled nebulae, with a
possible strong disc when a bi--stability limit
is crossed in the  equatorial region.
Thus, many features of the $\eta$ Carinae and 
LBV nebulae are  accounted for by the anisotropic
ejection from rotating stars.
               }

   \maketitle
%
%________________________________________________________________

\section{Introduction}

The HST picture of the  $\eta$ Carinae nebula (J. Morse and 
K. Davidson, STScI PRC96--23a) shows two big
symmetrical lobes and a disk-shaped skirt around the star
 (Davidson \& Humphreys  \cite{davidson97a}; Davidson et al. \cite
{davidson97b}). These authors find that the lobes 
 were created during the ``Great Eruption''of 1843,
 during which $\sim$ 1 to 3 $\mathrm{M_{\odot}}$ were ejected forming the Homunculus Nebula, with a total energy of $\sim 10^{49.5}$ ergs
(Humphreys \cite{humphreys99}).
The kinematics  confirms that it is  a bipolar
outflow (Nota \cite{nota99}), and the 
study of their limb darkening  shows that the lobes are hollow
(Davidson et al. \cite{davidson97a}).
The mass in the skirt 
amounts to 0.1 to 0.2 $\mathrm{M_{\odot}}$, it is still unclear 
whether it is resulting from the 1843 or later events. 
A massive cold torus of about 15 $\mathrm{M_{\odot}}$ 
has been found out in the equatorial plane (Morris et al. \cite {morris99}). 
Peanut--shaped nebulae and lobes are present around several 
other LBV stars (Nota \cite{nota97}).
%Recently, an unusual strong brightening of 
%$\eta$ Carinae has occured (Davidson et
%al \cite{davidson99}; Smith et al. \cite{smith00}).

Spectral variations
with a period of 5.5 yr have led to the suggestion that $\eta$ Carinae
is a binary system (Damineli \cite{dam96}; 
Lamers et al. \cite{lamers98}; Damineli et al. \cite{dam00}).
However,  the periodicity is not as strict as
claimed, with  shifts of 3 monthes between two
successive cycles, which casts some doubt
on the binary hypothesis (Smith et al. \cite{smith00})
   
Numerous  models for the $\eta$ Carinae nebula
have been proposed
(e.g. see Schulte--Ladbeck \cite{schulte97}; 
Hillier \cite{hillier97}). Several ones invoke
collisions of winds emitted 
at various  evolutionary stages, such as
the interaction of an isotropic fast wind with a previous slow
equatorially enhanced wind  (Kwok et al. 
\cite{kwok78}; Frank et al. \cite{frank95}).
The opposite scenario of  an aspherical
fast wind, expanding into a previously deposited slow spherical
 wind  has been studied by Frank et al.
(\cite{frank98}). This model recovers several features,
but not the equatorial skirt.
Langer (\cite{langer98}), Langer \& Heger (\cite{langheg98})
 consider that the LBV outbursts occur
when the outwards centrifugal and radiation forces
cancel gravity at the equator.  
Langer et al. (\cite{langer99}) develop a model where a slow, dense,
equatorial wind is followed by a fast and almost spherical wind
and this  leads to very representative
 simulations of the Homunculus Nebula.
The models  are using  the wind compressed disc model
 of Bjorkman and Cassinelli (\cite{bj93}). 

Owocki et al. (\cite{ow98}) have emphasized 
that the  account of the von Zeipel theorem (\cite{vonzeipel})
drastically modifies the wind distribution around a rotating star.
The properties of stellar winds ejected by rotating stars
 have been studied  (Maeder \cite{MIV}; Maeder \& Meynet
\cite{MMVI}), in particular the changes of the mass loss rates
 by rotation and the latitudinal distribution 
of the mass loss.
Here, we examine the distribution of the anisotropic wind 
around rotating star models corresponding to $\eta$ Carinae.
 The noticeable result is that 
there are bi--polar components and a disc naturally
arising in the  wind. 
Interactions 
of  the lobes and discs emitted at various phases during
the back and forth motions of an LBV in the HR diagram
may further shape the nebulae (cf. Langer et al.
\cite{langer99}).

% There are also  models 
%which invoke particular wind structures
%such as the Double--Bubble, the Bi--polar Cap, the Double--Flask or
%the Prussian Helmet
%(Hillier \cite{hillier97}). According to recent ESO observations 
%the  observed structure is well represented by the
%Double--Flask (Currie et al. \cite{Curr00}).

\section{The anisotropic mass loss}

\begin{figure}
\resizebox{\hsize}{!}{\includegraphics{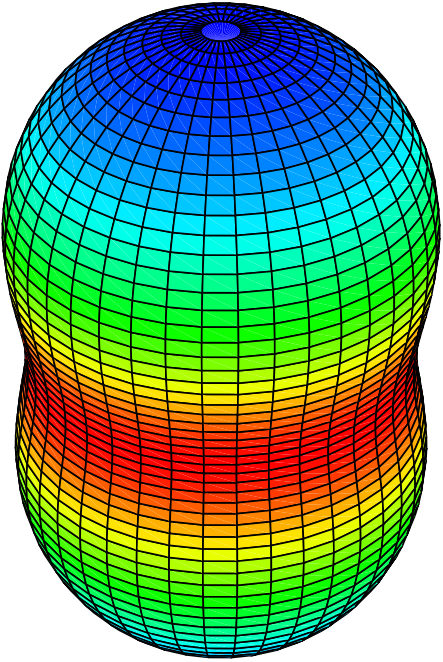}}
\caption{The mass fluxes around a rotating star like $\eta$ Carinae for 
a ratio of the angular velocity to the break-up angular velocity
$\omega = 0.80$, assuming a polar T$_{\mathrm{eff}} = 30000$ K.}
\label{Fig1}
\end{figure}

In rotating stars with high radiation
pressure, the surface shape is also described by the Roche model.
The total gravity is the sum of the gravitational, centrifugal and radiative accelerations:
$ \vec{g}_{\mathrm{tot}} = \vec{g}_{\mathrm{eff}} + \vec{g}_{\mathrm{rad}} = \vec{g}_{\mathrm{grav}} +
 \vec{g}_{\mathrm{rot}} + \vec{g}_{\mathrm{rad}} $.
The local flux at colatitude $\vartheta$ is proportional 
to the effective gravity $\vec{ g}_{\mathrm{eff} }$
according to the von Zeipel theorem 

   \begin{equation}
      \vec{F}(\vartheta) = -\frac{ L }{ 4\pi GM
                                     \left( 1-\frac{ \Omega^2 }
                                                   { 2\pi G \rho_m }
                                     \right)
                                     }\, 
                            \vec{ g}_{\mathrm{eff}}(\vartheta)  \, , 
   \end{equation}%eqn. 1

\noindent
where we ignore a small term ($\sim$ 1 \%) due to differential
rotation. The correct 
Eddington factor $\Gamma_{\Omega}(\vartheta)$
in a rotating star  depends on rotation (Maeder \& Meynet
\cite{MMVI}).

   \begin{equation}
     \Gamma_{\Omega}(\vartheta) = \frac{ \kappa(\vartheta) L }
                               { 4\pi cGM
                                  \left( 1-\frac{ \Omega^2 }
                                                { 2\pi G \rho_m }
                                  \right)
                               } \;,
       \end{equation}%eqn. 2

\noindent
where $\kappa(\vartheta)$ is the local opacity. 
One has

\begin{equation}
\vec{g}_{\mathrm{tot}} = \vec{g}_{\mathrm{eff}}
\left[ 1 - \Gamma_{\Omega}(\vartheta) \right] \; .
\end{equation}%eqn. 3

\noindent
The relation $\vec{g}_{\mathrm{tot}} = 0$ gives the critical velocity.
The  expression $v^2_{\rm{crit}} = 
\frac{GM}{R} (1-\Gamma) $ often used 
for massive stars  only applies
if the star is uniformly bright. If not, Eq. (3) has
two roots which need to be discussed (Maeder \& Meynet \cite{MMVI}). 
The result is that, due to the von Zeipel theorem, $v_{\rm{crit}}$ only
slowly tends towards zero for  high Eddington factor.

The distribution of the  mass loss by unit surface
at colatitude $\vartheta$ over the surface of a rotating
star is 

\begin{eqnarray}
\frac{\Delta\dot{M}(\vartheta)}{\Delta \sigma}
\simeq A \left[\frac{L(P)}{4\pi
GM  \left( 1 - \frac{\Omega^2}{2 \pi G \rho_{\rm{m}}}  \right)
  }\right]^{\frac{1}{\alpha}} \frac{g_\mathrm{eff}}
{[1 - \Gamma_{\Omega}]^{\frac{1}{\alpha}-1}}  \;,
\end{eqnarray}%eqn. 4

\noindent
where $\Gamma_{\Omega}$ is given by Eq. (2) for 
 electron scattering opacity. We have
 $A = \left(k\alpha\right)^{\frac{1}{\alpha}} \left(
\frac{1-\alpha}{\alpha}\right)^{\frac{1-\alpha}{\alpha}}$.
The parameter $\alpha$ is  a so-called force multiplier, 
which characterizes the intensities  and distribution of the strengths
of the spectral lines.
At some T$_{\mathrm{eff}}$, the ionisation equilibrium of the stellar
wind is changing rather abruptly and so does the opacity of the
plasma. Consequently,
the values of the force multipliers undergo rapid transitions,
e.g.   $\alpha = 0.54$ for log T$_{\mathrm{eff}}$ 
$\geq 4.342$, $\alpha = 0.30$ for log T$_{\mathrm{eff}}$
between 4.342 and 4.00 and $\alpha = 0.20$ for 
log T$_{\mathrm{eff}}$ $\leq 4.00$ (Lamers et al. \cite{lamers95}; Lamers \cite{lamers97}; see also  Kudritzki \& Puls \cite{kudr00}). Such 
transitions of the wind properties, which
 still have some  uncertainties, are  called bi--stability limits.
Since T$_{\mathrm{eff}}$ is
decreasing from pole to equator, one has the corresponding
variations of the force multipliers  and thus of 
the mass loss fluxes from pole to equator.

Eq. (4)  shows that two  main effects may
enhance the mass ejection. The first one is 
the higher gravity at the pole which makes it hotter, this is
the ``$g_\mathrm{eff}$--effect''. The second is
the ``$\kappa$--effect'', it is due to
the rapid growth of  the opacity  below a 
bi--stability limit. The result is a lower value
of  $\alpha$ in
the equatorial regions, if they become cool
enough; this  favours  strong equatorial ejections. 

\begin{figure}
\resizebox{\hsize}{!}{\includegraphics{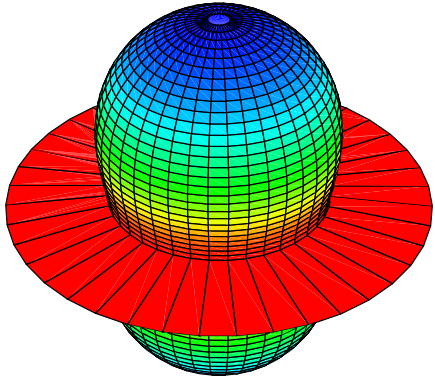}}
\caption{The mass flux around a rotating star with the mass
and luminosity of $\eta$ Carinae
for $\omega = 0.80$ and a polar
 T$_{\mathrm{eff}} = 25000$~K.}
\label{Fig2}
\end{figure}

We consider the case of 
$\eta$ Carinae, with a mass of $\sim$100 M$_{\odot}$,
a luminosity of  $ \sim 10^{6.5}$ 
$\mathrm{L_{\sun}}$, an Eddington factor 
$\Gamma$ = 0.823 for electron scattering opacity,
and a T$_{\mathrm{eff}} \simeq$  30000 K
or a few thousands degrees less at minimum
(Humphreys \cite{humphreys88};
Davidson et al. \cite{davidson97a}). Rotation is defined by
 $\omega$, the fraction of
the angular velocity at break--up, i.e.  $\omega^2 = \frac{\Omega^2 R^{3}_\mathrm{eb}}{GM}$. Fig. 1 shows the distribution of the 
mass fluxes for $\omega = 0.80$  and a reference polar
 T$_{\mathrm{eff}}$ =  30000 K.
We notice the prolate distribution with a peanut--shape,
which results from the g$_
{\mathrm{eff}}$--effect. For higher values of $\omega$, the
peanut-shape is more and more pronounced, while for lower $\omega$
we have a prolate spheroid.

When a bi--stability limit is crossed at some latitude
 on the rotating star, the $\dot{M}$--rates are strongly enhanced
at lower latitudes by the 
$\kappa$--effect. A crossing of the bi--stability limit
is favoured by fast rotation and a lower choice of the
reference  
T$_{\mathrm{eff}}$. Fig. 2 shows, as an example, the mass flux
for  $\omega = 0.8$ as in Fig. 1, but for 
T$_{\mathrm{eff}}$ = 25000 K. We notice the presence of a strong
equatorial ejection. The wider the latitude range with   T$_{\mathrm{eff}}$
below the bistability limit, the thicker the disk. 
%The global  mass loss rate increases
%strongly with rotation (cf. Maeder \& Meynet \cite{MMVI}).

\section{The shape of the nebulae}

The shape of the nebulae is given by the distribution of matter
at some time around the ejecting star. For purpose of simplicity, we 
consider here two simplified cases: a) a short time ejection,
b) a continuous ejection. 

\subsection{Short time ejection}

We assume that the geometry $r(t,\vartheta)$ at time $t$
of the shell ejected at $t_{0}$ is determined
 by the coasting velocity, 
while Eq.(4) determines the
amount of matter at a given position in the shell. Thus, one has
(cf. Maeder \cite{MIV})

\begin{eqnarray}
r(t,\vartheta) \;= \; v_{\infty}
(r,\vartheta) (t-t_{0})  \quad \quad \mathrm{with} \\[2mm] 
v_{\infty}(r,\vartheta) \simeq \frac{\alpha}{1-\alpha} \; g_{\mathrm{eff}}^
{\frac{1}{2}} \; r(\vartheta)^{\frac{1}{2}}
(1-\Gamma_{\Omega})^{\frac{1}{2}} \; .
\end{eqnarray}%equ.5

\noindent
Fig. 3 shows the distribution of matter for a short time ejection 
for a case   with $\omega =0.90$. If there is no bi--stability limit
crossed, this shape is the same for all kinds of stars 
with a given $\omega$ (cf. Eq. 6).
The nebula is hollow and  shows a 
peanut shape with an equatorial pinch. 
For lower
values of $\omega$,  the shape is a prolate spheroid; for higher 
$\omega$, the equatorial pinch is even more marked.
At a given $\omega$,
 the peanut shape is more pronounced for the $\dot{M}$ distribution
than for the $v_{\infty}$ distribution, since the first one depends 
on $g_{\mathrm{eff}}$, while the second one depends on
$g_{\mathrm{eff}}^{\frac{1}{2}}$. Such a
shape well corresponds to that of many LBV nebulae (Nota 
\cite{nota99}).

\begin{figure}
\resizebox{\hsize}{!}{\includegraphics{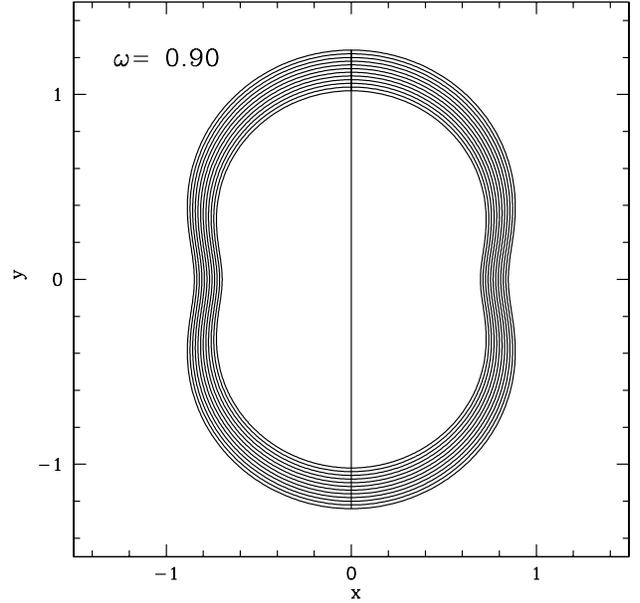}}
\caption{ Spatial distribution of the matter for a short ejection
around a star with $\omega =$ 0.90. The vertical line shows the direction
of the rotation axis.}
\label{Fig3}
\end{figure}

\subsection{Continuous ejection}

We may also assume the wind to be constant in time, 
and  for a long time. On each radial line,
the density decreases like $r^{-2}$. Thus, the lines 
$r(\vartheta)$ of constant densities are given by

\begin{eqnarray}
r(\vartheta) \; = \; \left[\frac{\rho(r_{0},\vartheta)}
{\rho(r,\vartheta)} \right]^{\frac{1}{2}} r_{0}(\vartheta)
\quad \mathrm{with} \\
\rho(r_{0},\vartheta) \simeq 
 \left(1-\alpha\right)^{\frac{1}{\alpha}}
\frac{ g_{\mathrm{eff}}^{\frac{1}{2}}}{r(\vartheta)^{\frac{1}{2}}
[1 - \Gamma_{\Omega}]^{\frac{1}{\alpha}-\frac{1}{2} }} \; .
\end{eqnarray}

\noindent
This model gives  centrally filled nebulae. Fig. 4 illustrates
the iso--density lines in the  case corresponding to  Fig. 1.
Since, the iso-density lines behave like $g_{\mathrm{eff}}^{\frac{1}{4}}$, 
the variations of $g_{\mathrm{eff}}$ have smaller effects and
the peanut shape is absent. We notice a 
strong  equatorial enhancement due to the 
lower $\alpha$ in these regions.
In real cases, if a sudden ejection is superposed on a more
continuous wind, we may have some compound of
the hollow peanut shell as in Fig. 3 and of the filled nebula
with a disc as in Fg. 4, well corresponding to 
 the various possible shapes of the  LBV
nebulae and $\eta$ Carinae.
 Interestingly enough, the group of the
B[e] stars (Zickgraf \cite{zickgraf}) shows a two--component 
stellar wind with a hot, highly ionized, fast wind at the poles
and a slow dense disk--like wind at the equator.

% Let us also point out that for a contracting star 
%keeping at break--up velocity, the rate of 
%mass loss will no longer  be given by the wind theory,
%but by the mechanical
%conditions of stability. 
%This situation will be analysed in another work.

\begin{figure}
\resizebox{\hsize}{!}{\includegraphics{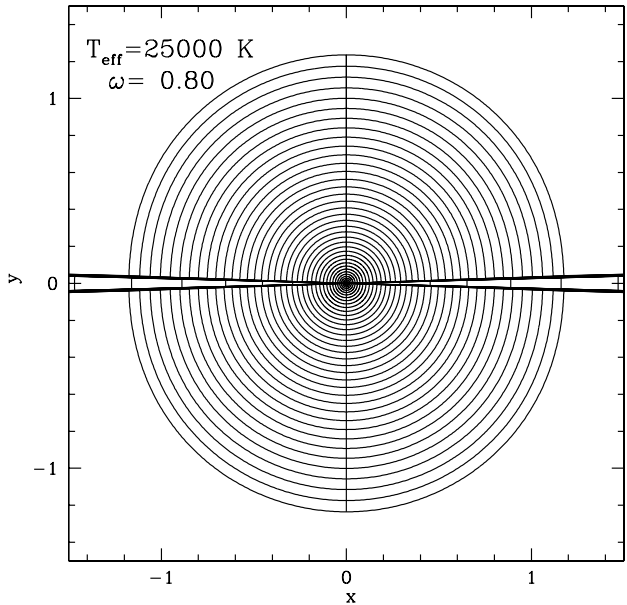}}
\caption{Iso--density lines for a continuous ejection
around a star with $\omega =$ 0.80. The density is strongly enhanced
in the equatorial plane.}
\label{Fig4}
\end{figure}

Figs. 3 and 4 only give two schematic 
examples of nebulae. Certainly, a kind of catalogue is to 
be made of all possible sorts of nebulae arising from
Eq. (4). This is a problem with many parameters:\\
-- Stellar rotation given by $\omega$,\\
-- Reference polar $T_{\mathrm{eff}}$ of the star,\\
-- Eddington factor $\Gamma$,\\
-- Opacities, bi--stability limits, opacity peaks, etc...\\
-- Particular history of mass ejection $\dot{M}(t)$.\\
-- Wind--wind interaction (cf. Langer et al. \cite{langer99}).\\
Certainly many interesting situations may occur. For example,
a B--star with a polar $T_{\mathrm{eff}}$ not so much above
the bi--stability limit could give rise to a substantial
equatorial ejection  and a disc formation, even if
the star is not at break--up !

Finally, we may wonder whether  LBV stars have fast rotation 
velocities. We see
two possibilities. a) If  due to a small
 magnetic field rotation keeps solid body 
(Langer \cite{langer98}), for most non--zero initial velocities
the star may   reach the break--up
velocity near the end of the MS. b) Models with
differential rotation also show that very massive stars may  reach the
so-called $\Omega \Gamma$--limit, i.e the  break--limit  function
of rotation and of the Eddington factor (Maeder \& Meynet \cite {MMVI}).
In particular, after the end of  the MS
phase, the stars  do a bluewards hook in the HR diagram; 
also after an outburst, they may
reach fast velocities as a result of the contraction of the outer layers.

\section{Conclusions}

Rotation produces 
strongly anisotropic mass loss 
in massive stars. 
Two schematic cases of winds have been considered.
The occurence of a
peanut--shaped nebula with  bi--polar lobes
and an equatorial disc  may
naturally result from the
the g$_{\mathrm{eff}}$ and  $\kappa$--effects.
Without particular wind interactions and collisions,
the wind  distribution
already possesses several important features of $\eta$ Carinae 
and of other LBV stars.

The critical question is   whether the polar lobes and  the skirt
in  $\eta$ Carinae result  directly 
from the anisotropic mass ejection (cf. Figs. 3 \& 4)
or whether they result from the interaction
of the  successive anisotropic winds.  
Whatever the answer, the anisotropic ejection with polar lobes and
an equatorial disc,  as shown here, 
is an essential part of the game.

%With some necessary reservations, 
%our preferred scenario is the following one.
%t
%rotation  reaches the $\Gamma$--limit 
%or the  $\Omega\Gamma$--limit when rotation is significant. 
%An outburst occurs (cf. geyser model, Maeder \cite{Mae92}). 
%The strong  mass loss is 
%essentially equatorial with a slow  dense wind,
%which might have produced the massive torus in  $\eta$ Carinae.
%As a result of the star readjustment,
%T$_{\mathrm{eff}}$ grows  quickly and
%$the star is keeping  close to break--up, with a faster  polar wind
%and a smaller equatorial ejection 
%($\alpha$ is larger for higher T$_{\mathrm{eff}}$ ),
%which may give rise to  an equatorial skirt. 

\begin{acknowledgements}
A.M.  expresses his best thanks  to Dr. Norbert Langer
and to Dr. Georges Meynet  for their
very helpful and constructive comments, and to Raoul Behrend
for his help in numerical modelling. 
\end{acknowledgements}

\end{document}